\pdfoutput=1

\documentclass[journal]{IEEEtran}

\usepackage[pdftex]{graphicx}
\usepackage{amsmath,amssymb}
\usepackage{cite}
\usepackage{placeins}

\graphicspath{{Figure/}}
\DeclareMathOperator*{\argmin}{arg\,min}
\DeclareMathOperator*{\argmax}{arg\,max}
\newcommand{\QG}{Q_{\mathrm G}}
\newcommand{\Vset}{\mathcal V}
\newcommand{\Bset}{\mathcal B}
\begin{document}

\title{Outage Analysis of Platoon-Head Selection for Vehicle Platooning in V2X Networks}

\author{Junhyeong~Kim%
\thanks{This work was supported by Institute of Information \& communications Technology Planning \& Evaluation (IITP) grant funded by the Korea government (MSIT) (No.RS-2024-00444230, Development of wireless technology for integrated sensing and communication).
\emph{(Corresponding author: Junhyeong Kim)}}%
\thanks{Junhyeong Kim is with the 6G Wireless Access System Research Section, Electronics and Telecommunications Research Institute (ETRI), Daejeon 34129, Republic of Korea. He is also with the Major in Information and Communication Engineering, ETRI School, University of Science and Technology (UST), Daejeon 34129, Republic of Korea (e-mail: jhkim41jf@etri.re.kr).}%
\thanks{This work has been submitted to the IEEE for possible publication. Copyright may be transferred without notice, after which this version may no longer be accessible.}%
}

\markboth{Preprint}%
{Kim: Outage Analysis of Platoon-Head Selection for V2X Platooning}

\maketitle

\begin{abstract}
This study analyzes the packet-outage performance of platoon-head selection for vehicle platooning in vehicle-to-everything (V2X) networks. A handover platoon head (H-PH) provides measurements for group handover, whereas a communication platoon head (C-PH) relays downlink packets. We derive serving-state-averaged outage expressions for three half-duplex decode-and-forward C-PH selection modes and for direct base-station (BS) groupcast under log-normal shadowing and Nakagami-$m$ fading. For block-level selection, an exact tail formulation reduces an $N$-dimensional shadowing expectation to a single integral. In a five-vehicle scenario with a 50-Mb/s target rate and the leading vehicle midway between two BSs, direct groupcast yields an outage probability of 0.269, whereas the block- and slot-level relay-selection modes yield 0.144 and 0.0459, respectively. Moving the H-PH from the leading to the middle vehicle reduces slot-level outage by 12.0\%. These results highlight the reliability benefits of H-PH placement and dynamic relay selection, as well as the potential gain of rapid intra-platoon channel state information sharing.
\end{abstract}


%**************************************************************************************
% Introduction
%**************************************************************************************
\section{Introduction}
\IEEEPARstart{V}{ehicle} platooning requires timely information exchange within a platoon and with roadside infrastructure. The 3GPP enhanced vehicle-to-everything (V2X) requirements include platooning services \cite{3GPP_TS22186_R19}. Packet loss and delay can degrade control stability and safety \cite{Zeng2019Joint,Oliveira2021Codesign}. Prior studies separately consider a platoon manager or leader for association and resource coordination \cite{Farzanullah2023Leader,Dong2022Manager} and a relay for packet delivery \cite{Li2020MmWaveRelay,Goncalves2023Relay}. Motivated by these distinct functions, this study considers two roles of platoon head (PH). The handover PH (H-PH) reports BS measurements, based on which the network determines a common serving base station (BS) for the platoon, while the communication PH (C-PH) relays downlink packets to the other platoon vehicles.

The reliability of these roles has mostly been studied through separate communication and mobility mechanisms. Repetition and relay selection improve vehicle-to-vehicle (V2V) groupcast reliability \cite{Kim2019Groupcast,Kuo2021Reliable}. Network-controlled group handover synchronizes the serving-cell change of a platoon \cite{vanDooren2018GroupHO,Chang2019Handover}, and learning-based handover policies have also been proposed \cite{Amaira2023Handover}.

Recent work addresses adjacent communication-design problems. Dynamic platoon-manager selection and finite-blocklength resource allocation were studied for intra-platoon groupcast \cite{Dong2022Manager}, while high-reliability resource allocation was optimized for direct infrastructure multicast \cite{Chen2023Multicast}. Multi-hop and beam-aware relay selection addresses mmWave range, blockage, and latency \cite{Li2020MmWaveRelay,Kim2023RelayBeam}. Cross-layer or network-aware relaying supports infrastructure-assisted and long platoons \cite{Goncalves2023Relay,Razzaghpour2024Mass}. These studies optimize a manager, resource allocation, or relay topology, but they do not connect the history-dependent serving-state distribution induced by one H-PH to the outage of a separately selected C-PH. They also do not compare fixed, block-level, and slot-level C-PH selection with direct BS groupcast in one analytical model.

To address these gaps, we analyze C-PH selection while accounting for the serving-BS distribution induced by H-PH-based group handover. This study makes three contributions. First, it derives exact packet-outage representations for fixed, block-level, and slot-level C-PH selection and direct groupcast under a unified model. Second, it incorporates the effect of H-PH-based group handover by mapping H-PH measurements to the network-controlled serving-state distribution and averaging every packet-outage expression over that state. Third, independence across candidate shadowing variables reduces block-level selection to a single exact one-dimensional integral, and the results quantify the effects of H-PH placement and C-PH channel state information (CSI)-update granularity, showing that rapid local CSI sharing for slot-level C-PH selection can substantially outperform direct BS groupcast. Monte Carlo (MC) simulations validate the analytical results for all communication modes.

%**************************************************************************************
% System Model
%**************************************************************************************
\section{System Model}
\begin{figure}[!t]
\centering
\includegraphics[width=\columnwidth]{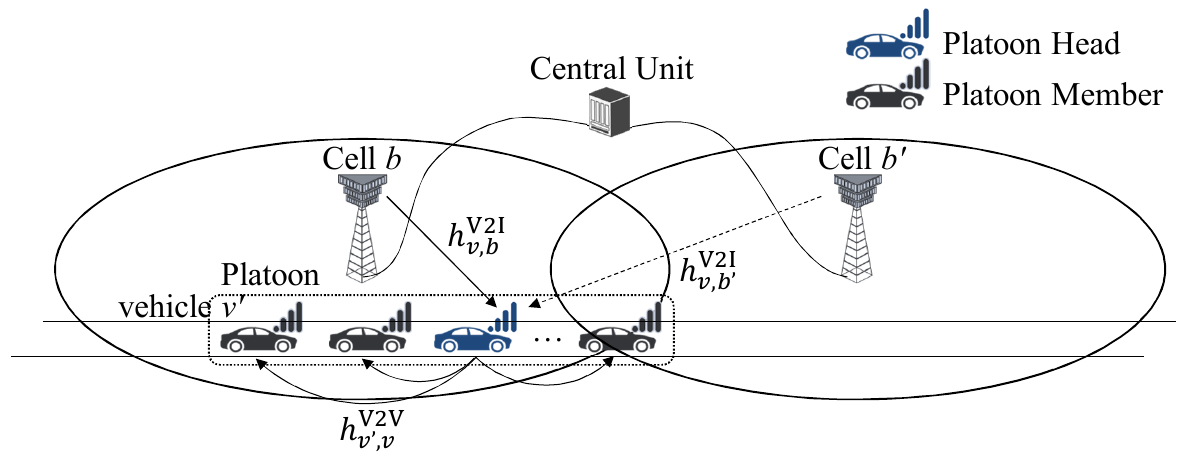}
\caption{Two-cell V2X platooning model. The H-PH provides measurements for network-controlled group handover, and the C-PH relays a BS packet. The drawing illustrates $h=c$; the analysis also allows $h\ne c$.}
\label{fig:system}
\end{figure}

As shown in Fig.~\ref{fig:system}, a platoon of $N$ vehicles $\Vset=\{1,\ldots,N\}$ travels past two BSs $\Bset=\{1,2\}$. Vehicle 1 leads, and adjacent vehicles are separated by $d_{\mathrm V}$. The H-PH $h$ reports BS measurements, based on which the network determines a common serving BS for the platoon. The C-PH $c$ receives a common packet over vehicle-to-infrastructure/network (V2I/N) and broadcasts it to every $u\in\Vset\setminus\{c\}$ over V2V. The two roles may be assigned to different vehicles.

In this study, only the H-PH performs and reports the intercell measurements used for group handover. Each C-PH candidate estimates its V2I/N and directed V2V links from reference signals. The candidates exchange these link metrics and the selected C-PH index only within the platoon. The serving BS transmits the common packet on the allocated platoon resource, so no C-PH metric or selected index is reported to the BS.

\subsection{Transmission Frame and Wireless Channel}
\begin{figure}[!t]
\centering
\includegraphics[width=0.94\columnwidth]{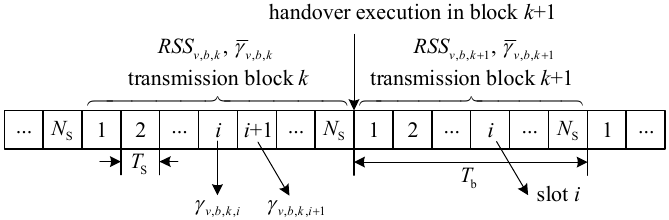}
\caption{Transmission frame structure. An RSS decision in block $k$ selects the serving BS for block $k+1$, and one slot carries one two-phase DF transmission attempt.}
\label{fig:frame}
\end{figure}

Fig.~\ref{fig:frame} illustrates the transmission frame structure. Each block has duration $T_{\mathrm b}=N_{\mathrm s}T_{\mathrm s}$ and contains $N_{\mathrm s}$ slots of duration $T_{\mathrm s}$. Shadowing and the serving BS are fixed within a block, while small-scale fading changes by slot. The network uses the H-PH measurement in block $k$ to determine the serving BS for block $k+1$, and handover completes before packet transmission. One packet transmission can use up to $N_{\mathrm{tx}}$ slots within a block, where $1\leq N_{\mathrm{tx}}\leq N_{\mathrm s}$. In each half-duplex decode-and-forward (DF) attempt, the C-PH receives during $\tau_{\mathrm I}T_{\mathrm s}$ and then broadcasts during $\tau_{\mathrm V}T_{\mathrm s}$, where $\tau_{\mathrm I}+\tau_{\mathrm V}=1$. Direct groupcast uses the full slot for one BS transmission to all vehicles.

The V2I/N and V2V phases share a common carrier resource pool but operate in orthogonal time intervals under half-duplex DF and thus do not interfere with each other. The V2I/N phase uses bandwidth $W_{\mathrm I}$, while the shorter-range V2V phase is allocated $W_{\mathrm V}\leq W_{\mathrm I}$. For link distance $d$, the large-scale power gain is $\kappa_0d^{-\eta}$. Let $a=\ln(10)/10$ and $\xi_{c,b,k}\sim\mathcal N(0,\sigma_{\mathrm{sh}}^2)$ denote the V2I/N shadowing on the link from BS $b$ to vehicle $c$ in
block $k$. The instantaneous signal-to-noise ratios (SNRs) in slot $i$ are
\begin{align}
\gamma^{\mathrm I}_{c,b,k,i}
 &=\overline\gamma^{\mathrm I}_{c,b,k}e^{a\xi_{c,b,k}}G^{\mathrm I}_{c,b,k,i},
&\overline\gamma^{\mathrm I}_{c,b,k}
 &=\frac{\kappa_0d_{c,b,k}^{-\eta}P_{\mathrm{Tx}}}{\sigma_{\mathrm I}^{2}},
\nonumber\\
\gamma^{\mathrm V}_{u,c,k,i}
 &=\overline\gamma^{\mathrm V}_{u,c}G^{\mathrm V}_{u,c,k,i},
&\overline\gamma^{\mathrm V}_{u,c}
 &=\frac{\kappa_0d_{u,c}^{-\eta}P_{\mathrm{Tx}}}{\sigma_{\mathrm V}^{2}}.
\label{eq:channel}
\end{align}
The unit-mean power gains obey
\begin{equation}
G^{\mathrm I}\sim\operatorname{Gamma}(m_{\mathrm I},1/m_{\mathrm I}),\qquad
G^{\mathrm V}\sim\operatorname{Gamma}(m_{\mathrm V},1/m_{\mathrm V}),
\label{eq:gamma_gain}
\end{equation}
where $m_{\mathrm I},m_{\mathrm V}\geq1/2$ \cite{Simon2005Fading}. In this study, shadowing is modeled only on the V2I/N links, while the V2V links have deterministic large-scale gains. V2I/N shadowing is independent across vehicles and blocks and is block-static. Conditioned on it, all V2I/N and directed V2V gains are mutually independent across candidates, receivers, links, and slots; reciprocal V2V directions are distinct links.

For end-to-end target rate $R$, the required SNR thresholds for the V2I/N and V2V phases are
\begin{equation}
\theta_{\mathrm I}=2^{R/(\tau_{\mathrm I}W_{\mathrm I})}-1,\qquad
\theta_{\mathrm V}=2^{R/(\tau_{\mathrm V}W_{\mathrm V})}-1.
\label{eq:hd_thresholds}
\end{equation}

\subsection{H-PH-Based Group Handover}
The received-signal-strength (RSS) measurement averages small-scale fading. Let $b$ and $b'$ denote the serving and target BS indices, respectively. The mean RSS at H-PH $h$ is $\mu_{b,h,k}=P_{\mathrm{Tx}}^{\mathrm{dBm}}+10\log_{10}(\kappa_0d_{h,b,k}^{-\eta})$. The network transitions from $b$ to $b'$ when the target RSS exceeds the serving RSS by hysteresis $H$. With independent equal-variance shadowing,
\begin{equation}
p_{h,k}^{b\rightarrow b'}=
\QG\!\left(\frac{H-\mu_{b',h,k}+\mu_{b,h,k}}
{\sqrt{2}\sigma_{\mathrm{sh}}}\right),
\label{eq:transition}
\end{equation}
where $\QG$ is the Gaussian tail function. For serving-state distribution $\boldsymbol\pi_{h,k}=[\pi_{h,k}^{(1)},\pi_{h,k}^{(2)}]$,
\begin{align}
\pi_{h,k+1}^{(1)}
 &=\pi_{h,k}^{(1)}(1-p_{h,k}^{1\to2})
 +\pi_{h,k}^{(2)}p_{h,k}^{2\to1},
\nonumber\\
\pi_{h,k+1}^{(2)}
 &=\pi_{h,k}^{(1)}p_{h,k}^{1\to2}
 +\pi_{h,k}^{(2)}(1-p_{h,k}^{2\to1}),
\quad\boldsymbol\pi_{h,0}=[1,0].
\label{eq:state}
\end{align}
The analysis and simulation evaluate every $h\in\Vset$ to quantify how H-PH placement changes the serving-state distribution and packet reliability.

%**************************************************************************************
% C-PH Selection and Packet Outage Analysis
%**************************************************************************************
\section{C-PH Selection and Packet Outage Analysis}
We consider three vehicular platoon communication (VPC) modes with fixed, block-level, and slot-level C-PH selection, denoted by VPC~1, VPC~2, and VPC~3, respectively. In all three modes, a packet transmission attempt is successful only if its V2I/N reception and V2V broadcast phases both succeed in the same slot.

The three VPC modes differ in the channel information available for C-PH selection. VPC 1 uses no online channel information and employs a configured fixed relay. VPC 2 assumes block-level large-scale CSI, including path loss and block-static shadowing. VPC 3 assumes perfect instantaneous CSI, including both large- and small-scale channel gains, and therefore serves as an ideal full-CSI benchmark.

We define the regularized upper incomplete gamma function $\mathcal Q(m,z)=\Gamma(m,z)/\Gamma(m)$. Under the mutual-independence assumptions above, the V2I/N success, all-member V2V success, and end-to-end failure probabilities for candidate $c$, BS $b$, and shadowing $\xi$ are
\begin{align}
S^{\mathrm I}_{c,b,k}(\xi)
 &=\mathcal Q\!\left(m_{\mathrm I},
 \frac{m_{\mathrm I}\theta_{\mathrm I}}
 {\overline\gamma^{\mathrm I}_{c,b,k}e^{a\xi}}\right),
\nonumber\\
S^{\mathrm V}_{c}
 &=\prod_{u\in\Vset\setminus\{c\}}
 \mathcal Q\!\left(m_{\mathrm V},
 \frac{m_{\mathrm V}\theta_{\mathrm V}}
 {\overline\gamma^{\mathrm V}_{u,c}}\right),
\nonumber\\
q_{c,b,k}(\xi)&=1-S^{\mathrm I}_{c,b,k}(\xi)S^{\mathrm V}_{c}.
\label{eq:q}
\end{align}
For $m_{\mathrm I}=m_{\mathrm V}=1$, $\mathcal Q(1,z)=e^{-z}$ gives the Rayleigh special case. A complete DF transmission attempt comprises V2I/N reception followed by V2V broadcast. Failed attempts are discarded without combining, so the conditional packet outage for candidate $c$ is $q_{c,b,k}(\xi)^{N_{\mathrm{tx}}}$.

The shadowing-averaged outage of a C-PH used in every transmission slot is
\begin{align}
\Psi_{c,b,k}
 &=\frac{1}{\sqrt{2\pi}\sigma_{\mathrm{sh}}}
 \int_{-\infty}^{\infty}q_{c,b,k}(\xi)^{N_{\mathrm{tx}}}
 e^{-\xi^2/(2\sigma_{\mathrm{sh}}^2)}d\xi
\nonumber\\
 &=\frac{1}{\sqrt\pi}\int_{-\infty}^{\infty}e^{-x^2}
 q_{c,b,k}(\sqrt2\sigma_{\mathrm{sh}}x)^{N_{\mathrm{tx}}}dx
\nonumber\\
 &\simeq\frac{1}{\sqrt\pi}\sum_{r=1}^{M}w_r
 q_{c,b,k}(\sqrt2\sigma_{\mathrm{sh}}x_r)^{N_{\mathrm{tx}}}.
\label{eq:psi}
\end{align}
The second line follows from the substitution $x=\xi/(\sqrt{2}\sigma_{\mathrm{sh}})$. The last line applies an $M$-point Gauss--Hermite rule, whose $r$th node--weight pair is $(x_r,w_r)$, to the Gaussian-weighted integral. Increasing $M$ refines this one-dimensional approximation without introducing a multidimensional shadowing grid.

\subsection{Fixed C-PH (VPC 1)}
VPC 1 assigns one C-PH $c_{\mathrm F}$ by configuration and keeps that vehicle for every block and transmission slot. Its packet outage is
\begin{equation}
P_{h,k}^{\mathrm{VPC1}}=\sum_{b\in\Bset}\pi_{h,k}^{(b)}\Psi_{c_{\mathrm F},b,k}.
\label{eq:vpc1}
\end{equation}
The serving-state probability $\pi_{h,k}^{(b)}$ is determined by H-PH $h$, while $\Psi_{c_{\mathrm F},b,k}$ describes the reliability of fixed C-PH $c_{\mathrm F}$.

\subsection{Block-Level C-PH Selection (VPC 2)}
VPC 2 selects one C-PH from block-level link metrics and retains it for every transmission slot of the packet. For fixed $b$ and $k$, define $Y_c=q_{c,b,k}(\xi_c)$ and $Z=\min_{c\in\Vset}Y_c$. Thus $c_k^*=\argmin_cY_c$ and the selected candidate has conditional outage $Z^{N_{\mathrm{tx}}}$. Because $0\leq Z\leq1$, the tail-integral identity gives
\begin{align}
\mathbb E_Z[Z^{N_{\mathrm{tx}}}]
 &=\int_0^1\Pr(Z^{N_{\mathrm{tx}}}>y)\,dy
\nonumber\\
 &=N_{\mathrm{tx}}\int_0^1r^{N_{\mathrm{tx}}-1}
 \Pr(Z>r)\,dr
\nonumber\\
 &=N_{\mathrm{tx}}\int_0^1r^{N_{\mathrm{tx}}-1}
 \prod_{c\in\Vset}\Pr(Y_c>r)\,dr.
\label{eq:tail_identity}
\end{align}
The first equality expresses the mean of the bounded nonnegative variable $Z^{N_{\mathrm{tx}}}$ as the area under its tail probability. The substitution $y=r^{N_{\mathrm{tx}}}$ gives the second equality. Finally, independent candidate shadowing factorizes $\Pr(Z>r)=\Pr(Y_c>r,\,\forall c)$ into the product in the last line. This conversion is useful because it removes direct integration over the $N$-dimensional vector $\boldsymbol\xi$.

Because $\xi$ raises the mean V2I/N SNR in \eqref{eq:channel}, $q_{c,b,k}$ decreases monotonically with $\xi$. Define $\overline F_{c,b,k}(r)=\Pr(Y_c>r)$. For $1-S_c^{\mathrm V}<r<1$, inversion of \eqref{eq:q} yields
\begin{equation}
\begin{aligned}
\overline F_{c,b,k}(r)&=\Phi\!\left(\frac{\zeta_{c,b,k}(r)}
{a\sigma_{\mathrm{sh}}}\right),\\
\zeta_{c,b,k}(r)&=\ln\!\left[
\frac{m_{\mathrm I}\theta_{\mathrm I}}
{\overline\gamma^{\mathrm I}_{c,b,k}\,
\mathcal Q^{-1}_{2}(m_{\mathrm I},(1-r)/S_c^{\mathrm V})}\right],
\end{aligned}
\label{eq:survival}
\end{equation}
where $\Phi$ is the standard Gaussian cumulative distribution function (CDF) and $\mathcal Q^{-1}_{2}$ inverts the second argument. The survival function equals one below $1-S_c^{\mathrm V}$ and zero at $r\geq1$. Consequently,
\begin{align}
\overline P_{b,k}^{\mathrm{VPC2}}
 &=N_{\mathrm{tx}}\int_0^1r^{N_{\mathrm{tx}}-1}
 \prod_{c\in\mathcal V}\overline F_{c,b,k}(r)\,dr,
\nonumber\\
P_{h,k}^{\mathrm{VPC2}}
 &=\sum_{b\in\Bset}\pi_{h,k}^{(b)}\overline P_{b,k}^{\mathrm{VPC2}}.
\label{eq:vpc2}
\end{align}
This representation replaces an $M^N$ tensor quadrature by one adaptive integral whose integrand contains $N$ scalar survival functions. Its per-position cost therefore grows linearly rather than exponentially with $N$.

\subsection{Slot-Level C-PH Selection (VPC 3)}
For an instantaneous channel realization, define candidate $c$'s normalized decoding margin and the VPC 3 selection rule as
\begin{equation}
\begin{aligned}
\Lambda_{c,b,k,i}&=\min\!\left\{\frac{\gamma^{\mathrm I}_{c,b,k,i}}{\theta_{\mathrm I}},
\min_{u\in\Vset\setminus\{c\}}\frac{\gamma^{\mathrm V}_{u,c,k,i}}{\theta_{\mathrm V}}\right\},\\
c_{b,k,i}^{\star}&\in\argmax_{c\in\Vset}\Lambda_{c,b,k,i}.
\end{aligned}
\label{eq:vpc3_selection}
\end{equation}
Conditioned on $\xi_c$, candidate $c$ fails iff $\Lambda_{c,b,k,i}<1$, with probability $q_{c,b,k}(\xi_c)$. The selected attempt therefore fails iff all candidates fail. With independent slot fading and block-static $\boldsymbol\xi$,
\begin{align}
\overline P_{b,k}^{\mathrm{VPC3}}
 &=\mathbb E_{\boldsymbol\xi}\!\left[
 \left(\prod_{c\in\Vset}q_{c,b,k}(\xi_c)\right)^{N_{\mathrm{tx}}}
 \right]
\nonumber\\
 &=\mathbb E_{\boldsymbol\xi}\!\left[
 \prod_{c\in\Vset}q_{c,b,k}(\xi_c)^{N_{\mathrm{tx}}}\right]
 =\prod_{c\in\Vset}\Psi_{c,b,k},
\nonumber\\
P_{h,k}^{\mathrm{VPC3}}
 &=\sum_{b\in\Bset}\pi_{h,k}^{(b)}
 \overline P_{b,k}^{\mathrm{VPC3}}.
\label{eq:vpc3}
\end{align}
Independent slot fading gives the $N_{\mathrm{tx}}$th power, while independent candidate shadowing and directed gains with \eqref{eq:psi} give the final product. Correlation would invalidate this factorization. Thus $P^{\mathrm{VPC3}}\leq P^{\mathrm{VPC2}}\leq P^{\mathrm{VPC1}}$; VPC 3 is an ideal full-CSI benchmark, whereas VPC 2 uses block-level large-scale metrics.

\subsection{Direct BS Groupcast}
As a nonrelay benchmark, the serving BS uses the full slot to groupcast directly to all vehicles at rate $R$, giving $\theta_{\mathrm G}=2^{R/W_{\mathrm I}}-1$. Let
$S^{\mathrm G}_{u,b,k}(\xi_u)=\mathcal Q(m_{\mathrm I},m_{\mathrm I}\theta_{\mathrm G}/(\overline\gamma^{\mathrm I}_{u,b,k}e^{a\xi_u}))$ and $M_{u,b,k}^{(j)}=\mathbb E_{\xi_u}[(S^{\mathrm G}_{u,b,k})^j]$. By definition, $M_{u,b,k}^{(0)}=1$ because $(S^{\mathrm G}_{u,b,k})^0=1$. This supplies the $j=0$ term of the binomial expansion. The conditional all-vehicle success and packet-outage probabilities are
\begin{align}
A_{b,k}(\boldsymbol\xi)
 &=\prod_{u\in\Vset}S^{\mathrm G}_{u,b,k}(\xi_u),
&P_{b,k}^{\mathrm G}\mid\boldsymbol\xi
 &=[1-A_{b,k}(\boldsymbol\xi)]^{N_{\mathrm{tx}}}.
\label{eq:groupcast_conditional}
\end{align}
The shadowing vector is constant over the packet's transmission slots, whereas the Nakagami gains are independently realized in each slot. Therefore, the expectation over $\boldsymbol\xi$ must be taken after the $N_{\mathrm{tx}}$th power. Applying the binomial theorem before averaging factorizes the multidimensional expectation:
\begin{equation}
P_{h,k}^{\mathrm G}=\sum_{b\in\Bset}\pi_{h,k}^{(b)}
\sum_{j=0}^{N_{\mathrm{tx}}}(-1)^j\binom{N_{\mathrm{tx}}}{j}
\prod_{u\in\Vset}M_{u,b,k}^{(j)},
\label{eq:groupcast}
\end{equation}
where each moment requires only the one-dimensional quadrature in \eqref{eq:psi}.

\begin{figure*}[!t]
\centering
\includegraphics[width=0.90\textwidth]{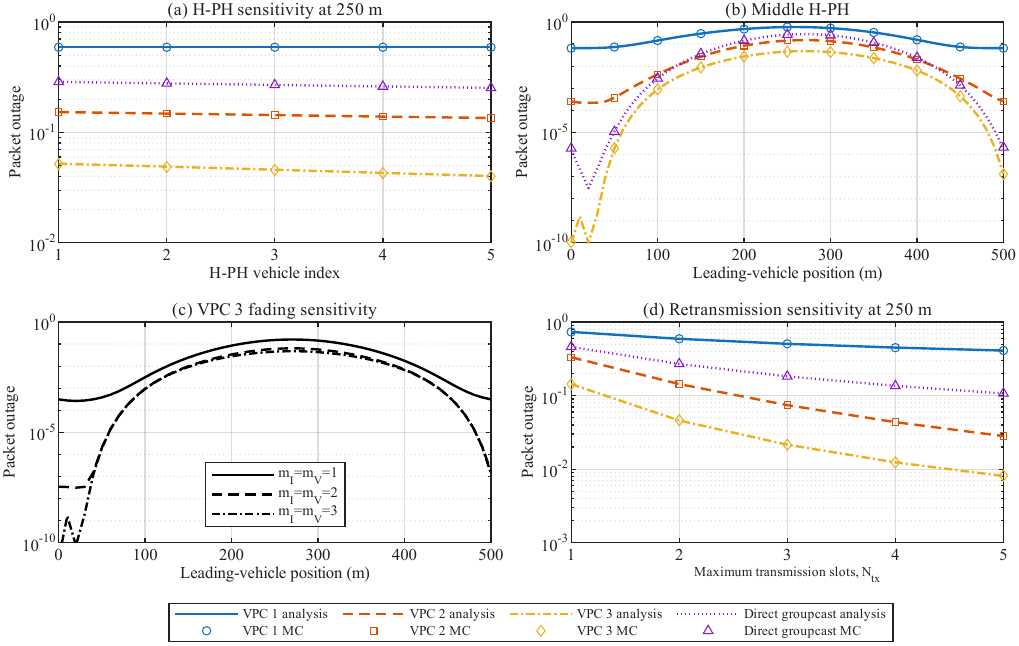}
\caption{Outage of half-duplex DF and direct groupcast: (a) H-PH candidates at 250~m; (b) position sweep with the middle H-PH; (c) analytical VPC 3 fading sensitivity; and (d) $N_{\mathrm{tx}}$ sensitivity at 250~m. Analysis and MC are separate legend entries in (a), (b), and (d); black lines distinguish fading cases in (c). MC position markers appear every 50~m, and clipped values are shown at the lower axis bound.}
\label{fig:results}
\end{figure*}

%**************************************************************************************
% Numerical Results
%**************************************************************************************
\section{Numerical Results}
Two BSs are located at longitudinal positions 0 and 500~m, 5~m from the road, and 10~m high. Five vehicles with antennas mounted at a height of 1.5 m and 10-m inter-vehicle spacing travel at $v=60$ km/h, and VPC 1 fixes $c_{\mathrm F}=1$. With $T_{\mathrm b}=10$~ms, the serving-state recursion in \eqref{eq:state} advances every $vT_{\mathrm b}=1/6$~m. Outage is evaluated every $L=60$ blocks, giving spatial interval $\Delta x=LvT_{\mathrm b}=10$~m. Table~\ref{tab:parameters} lists the remaining parameters. The line-of-sight reference model uses $\eta=2$, $\kappa_0=(c_{\mathrm{light}}/(4\pi f_c))^2$, where $c_{\mathrm{light}}$ is the speed of light, and 0-dBi antenna gains.

The shared carrier resource pool assigns $W_{\mathrm I}=20$~MHz to V2I/N and $W_{\mathrm V}=10$~MHz to the shorter-range V2V phase, which represents the generally smaller allocation available to the sidelink. Direct groupcast uses $W_{\mathrm I}$ for the full slot, whereas each DF attempt uses $W_{\mathrm I}$ and $W_{\mathrm V}$ only during its respective half-slot. Thus the direct benchmark is deliberately given a larger time--bandwidth allocation than VPCs 1--3.

\begin{table}[!t]
\caption{Baseline Numerical Parameters}
\label{tab:parameters}
\centering
\footnotesize
\renewcommand{\arraystretch}{1.02}
\begin{tabular}{@{}ll@{}}
\hline
Parameter & Value \\
\hline
Carrier frequency, transmit power & 30~GHz, 30~dBm \\
V2I/N, V2V bandwidth & 20, 10~MHz \\
End-to-end target rate $R$ & 50~Mb/s \\
DF time fractions $(\tau_{\mathrm I},\tau_{\mathrm V})$ & $(0.5,0.5)$ \\
Maximum transmission slots $N_{\mathrm{tx}}$ & 2 \\
Block duration $T_{\mathrm b}$, speed $v$ & 10~ms, 60~km/h \\
Sampling stride $L$, interval $\Delta x$ & 60 blocks, 10~m \\
Shadowing deviation, hysteresis & 4, 2~dB \\
Noise figure, noise density & 7~dB, $-174$~dBm/Hz \\
Platoon size, spacing & 5 vehicles, 10~m \\
Path-loss exponent, antenna gain & 2, 0~dBi \\
Main fading shapes $(m_{\mathrm I},m_{\mathrm V})$ & $(2,3)$ \\
Quadrature order $M$, MC samples/BS $N_{\mathrm{MC}}$ & 20, $2\times10^5$ \\
\hline
\end{tabular}
\end{table}

The MC simulations are stratified by serving BS. For each serving-BS state, $N_{\mathrm{MC}}$ shadowing vectors are generated, and the BS-conditioned outage is evaluated by conditionally averaging Nakagami-$m$ fading through \eqref{eq:q}. The conditional estimates are then weighted by $\pi_{h,k}^{(b)}$, ensuring that even rarely visited serving states are represented.

\subsection{H-PH Selection}
Fig.~\ref{fig:results}(a) evaluates all H-PH candidates $h=1,\ldots,5$ at 250~m. Changing $h$ from vehicle 1 to 5 reduces VPC 2, VPC 3, and direct-groupcast outage by 11.7\%, 22.7\%, and 11.8\%, reaching 0.1351, 0.04030, and 0.2531. The middle H-PH gives a VPC 3 outage of 0.04591, 12.0\% below the leading-H-PH result. VPC 1 remains at 0.5906 because its two BS-conditioned outages are equal at this symmetric geometry. More generally, H-PH placement changes the accumulated serving-state probabilities in \eqref{eq:state}.

\subsection{Relay Selection Versus Direct Groupcast}
Fig.~\ref{fig:results}(b) compares the four communication modes with the middle H-PH. At 250~m, direct groupcast has a lower outage than VPC~1, 0.2691 versus 0.5906, because it avoids the half-duplex V2V phase and receives the larger time--bandwidth allocation described above. Block-level VPC 2 and slot-level VPC 3 nevertheless attain 0.1435 and 0.04591, respectively, which are 46.7\% and 82.9\% below direct groupcast. Dynamic C-PH selection therefore compensates for the half-duplex rate penalty.

\subsection{Fading, Retransmissions, and Practical Considerations}
Fig.~\ref{fig:results}(c) shows that the VPC 3 outage decreases as $m_{\mathrm I}$ and $m_{\mathrm V}$ increase, reflecting the reduced severity of fading. The case $m_{\mathrm I}=m_{\mathrm V}=1$ corresponds to Rayleigh fading. In Fig.~\ref{fig:results}(d), increasing $N_{\mathrm{tx}}$ from one to two reduces VPC 3 outage from 0.1433 to 0.04591, or by 68.0\%, while direct-groupcast outage falls by 41.5\% to 0.2691. At $N_{\mathrm{tx}}=5$, VPC 3 reaches 0.00809 and is 92.4\% below direct groupcast. Additional slots provide diminishing returns.

The VPC 3 selection gain relies on the assumed independence of candidate-specific V2I/N shadowing and directed V2V fading gains. Common blockage or correlated links would correlate candidate failures and reduce the selection-diversity gain, making block-level VPC 2 more practical when instantaneous CSI acquisition is costly. The analytical results show close agreement with the MC simulations in Fig.~\ref{fig:results}(a), (b), and (d), with most MC markers overlapping the corresponding analytical curves at the plotting resolution.

%**************************************************************************************
% Conclusion
%**************************************************************************************
\section{Conclusion}
We analyzed packet outage for three C-PH selection modes and direct groupcast in V2X vehicle platooning. H-PH-based group handover was incorporated through the H-PH-dependent serving-BS distribution, and block-level selection was reduced to a single one-dimensional integral. Stratified MC simulations showed close agreement with the analysis. Under the baseline setting, direct groupcast outperformed the fixed vehicle relay, whereas dynamic C-PH selection achieved lower outage despite half-duplex operation. H-PH placement and the retransmission budget also materially affected reliability. Because slot-level selection is an ideal full-CSI benchmark, future work should account for correlated blockage, CSI acquisition and exchange overhead, and handover interruption and control latency.

\bibliographystyle{IEEEtran}
\bibliography{Source/VP_V2X_refs}

\end{document}